\documentclass[letterpaper, 10 pt, conference]{ieeeconf}  % Comment this line out if you need a4paper

\IEEEoverridecommandlockouts                              % This command is only needed if 
\usepackage{graphicx}
\usepackage{url}
\usepackage{hyperref}
\usepackage[utf8]{inputenc}
\usepackage{hyperref}
\hypersetup{unicode=true}
\usepackage{algorithm}
\usepackage{algorithmic}
\usepackage{booktabs}
\usepackage{tikz}
\usetikzlibrary{shapes, arrows, shadows, positioning}
\usepackage{pgfplots}

\usepgfplotslibrary{fillbetween}
\usepackage{colortbl}
\usepackage{ctable}
\usepackage{amsmath}
\usepackage{amssymb}

\title{\LARGE \bf
Feedback dynamics in Politics:\\The interplay between sentiment and engagement*
}

\author{Simone Formentin %<-this stops a space
\thanks{*This paper is partially supported by FAIR (Future Artificial Intelligence Research) project, funded by the NextGenerationEU program within the PNRR-PE-AI scheme (M4C2, Investment 1.3, Line on Artificial Intelligence), by the Italian Ministry of Enterprises and Made in Italy in the framework of the project 4DDS (4D Drone Swarms) under grant no. F/310097/01-04/X56 and by the PRIN PNRR project  P2022NB77E “A data-driven cooperative framework for the management of distributed energy and water resources” (CUP: D53D23016100001), funded by the NextGeneration EU program. It has been also partly supported by the project “European Lighthouse to Manifest Trustworthy and Green AI” (ENFIELD), which have received funding from the European Union’s Horizon Europe research and innovation program under grant agreement No. 101120657.}% <-this % stops a space
\thanks{Simone Formentin is with Dipartimento di Elettronica, Informazione e Bioingegneria, Politecnico di Milano, via G. Ponzio 34/5 - 20133 Milan, Italy. Email to:
        {\tt\small simone.formentin@polimi.it}.}%
}

\begin{document}

\maketitle
\thispagestyle{empty}
\pagestyle{empty}

%%%%%%%%%%%%%%%%%%%%%%%%%%%%%%%%%%%%%%%%%%%%%%%%%%%%%%%%%%%%%%%%%%%%%%%%%%%%%%%%
%\begin{abstract}
%In this work, we experimentally investigate the hypothesis that politicians adapt the sentiment of their online messages {\it in response to engagement signals}, thus closing a feedback loop between public reaction and future communication. To this purpose, we analyze a dataset of over 1.5 million tweets from Members of Parliament in the United Kingdom, Spain, and Greece during 2021, using a multilingual sentiment classifier. A simple yet interpretable linear predictor is proposed to capture sentiment dynamics at both aggregate and individual levels. Our analysis clearly indicate that future sentiment is dynamically influenced by the engagement levels of positive and negative messages. Moreover, by studying the spread of the learnt model coefficients, we identify systematic variations across political roles: opposition members are more reactive to negative engagement, whereas government officials respond more to positive signals. These findings provide a data-driven explanation for the proven amplification of negativity in online political discourse, and demonstrate how feedback principles can illuminate complex socio-political behavior.
%\end{abstract}

\begin{abstract}
We investigate feedback mechanisms in political communication by testing whether politicians adapt the sentiment of their messages \textit{in response to public engagement}. Using over 1.5 million tweets from Members of Parliament in the United Kingdom, Spain, and Greece during 2021, we identify sentiment dynamics through a simple yet interpretable linear model. The analysis reveals a closed-loop behavior: engagement with positive and negative messages influences the sentiment of subsequent posts. Moreover, the learned coefficients highlight systematic differences across political roles: opposition members are more reactive to negative engagement, whereas government officials respond more to positive signals. These results provide a quantitative, control-oriented view of behavioral adaptation in online politics, showing how feedback principles can explain the self-reinforcing dynamics that emerge in social media discourse.
\end{abstract}

%%%%%%%%%%%%%%%%%%%%%%%%%%%%%%%%%%%%%%%%%%%%%%%%%%%%%%%%%%%%%%%%%%%%%%%%%%%%%%%%
\section{INTRODUCTION}
The dynamics of modern political communication are increasingly mediated through social media platforms~\cite{perloff2021dynamics,schloz2024double,sianturi2024impact,magdaci2022modeling}. On channels such as X (formerly Twitter), politicians broadcast messages daily and receive immediate public feedback in the form of likes, shares, and replies. In this context, {\it sentiment analysis} has become a central tool for quantifying the impact of political communication, where sentiment denotes the polarity or affective orientation extracted from textual content through natural language processing methods. Prior studies have shown that negatively charged political messages tend to propagate more widely than positive ones~\cite{antypas2023negativity}, and that sentiment analysis can even serve as a predictor of election outcomes~\cite{alvi2023twitter}.

Beyond content analysis, several recent works have interpreted online political communication through the lens of {\it feedback systems theory}. The presence of feedback loops in media has long been recognized in communication research:~\cite{trilling2024communicative} frame digital discourse as interconnected loops where audience actions influence algorithmic outputs and future content exposure. Social influence has been modeled as an internal feedback-control process~\cite{weiss2024social}, while other works apply control-theoretic concepts directly to social dynamics. For instance,~\cite{sprenger2024control} design controllers to maximize engagement in a Friedkin--Johnsen opinion network, and~\cite{debuse2023opinion} introduce a strategic agent that adaptively reshapes opinions to reduce echo chambers. Related contributions include hypergradient-based recommender design for network interventions~\cite{kuhne2025optimizing} and studies on social pressure and stochastic opinion evolution~\cite{jadbabaie2022inference,tang2025stochastic}.

Despite this growing literature, most of these studies conceptualize feedback mechanisms from theoretical, psychological, or algorithmic perspectives. Yet, empirical evidence on whether and how human agents - particularly elected politicians - adapt their communication based on feedback remains missing. In other words, while previous works model the media environment as a feedback system, it is still unclear whether politicians themselves act as adaptive components within that loop.

This paper addresses this gap by investigating whether politicians adjust the sentiment of their future tweets in response to public engagement signals, thus {\it closing the loop}. Building on~\cite{antypas2023negativity}, we model the dynamic relationship between tweet sentiment and engagement through a simple autoregressive framework. The analysis is conducted on the same large-scale dataset of tweets from Members of Parliament (MPs) in the United Kingdom, Spain, and Greece during 2021, using the multilingual sentiment classifier XLM-T-Sent for consistency. 

Despite its simplicity, the proposed approach provides novel, data-driven insights into the adaptive behavior of politicians on social media. By estimating interpretable linear models, we quantify each politician's sensitivity to public engagement and reveal systematic variations across political roles and parties. The findings offer an empirical explanation for the self-reinforcing dynamics observed in~\cite{antypas2023negativity}.

In contrast to existing works~\cite{trilling2024communicative}--\cite{tang2025stochastic}, which focus on the structure or control of social feedback systems, this study provides {\it the first empirical evidence that real politicians modulate the emotional tone of their communication based on audience feedback}. Our linear modeling framework, applied to multilingual real-world data, offers a quantitative characterization of feedback adaptation in political communication.

{In summary, this paper contributes the following:}
\begin{itemize}
  \item It provides the first empirical evidence that elected politicians adjust the sentiment of their social media communication in response to public engagement.
  \item It introduces an interpretable model that captures this feedback loop and reveals systematic behavioral differences across political parties and roles.
  \item It complements the empirical analysis with a theoretical exploration of the closed-loop sentiment-engagement dynamics, highlighting conditions under which political discourse may stabilize or polarize.
\end{itemize}

The remainder of the paper is organized as follows. Section~\ref{sec:setup} describes the dataset and sentiment analysis methodology. Section~\ref{sec:feedback} introduces the feedback model linking sentiment and engagement and presents aggregate results. Section~\ref{sec:individuals} extends the analysis to individual politicians, highlighting cross-party and cross-role patterns. Section~\ref{sec:theory} discusses theoretical implications of the closed-loop dynamics, and Section~\ref{sec:conclusions} concludes the paper.

%%%%%%%%%%%%%%%%%%%%%%%%%%%%%%%%%%%%%%%%%%%%%%%%%%%%%%%%%%%%%%%%%%%%%%%%%%%%%%%%%%%%%%%
\section{EXPERIMENTAL SETUP}\label{sec:setup}
We employed a comprehensive dataset of tweets authored by Members of Parliament (MPs) from three European countries — the United Kingdom (UK), Spain, and Greece — covering the period from {January to December 2021}. This timeframe was selected to avoid distortions due to national elections (held in 2019 for all three countries) and to capture political communication during a shared exogenous shock: the COVID-19 pandemic.

The dataset comprises a total of {1,588,970 original tweets} from {2,213 MPs}, with retweets excluded to focus on the primary virality metric: {retweet count}. Data collection was performed via the Twitter API by the authors of~\cite{antypas2023negativity}, whom we gratefully acknowledge. Only tweets with substantive text content were retained.

Sentiment analysis was conducted using the {XLM-T-Sent} model, a multilingual transformer-based classifier fine-tuned on in-domain data. This model was chosen for its robustness across multiple languages, including lower-resource ones such as Catalan, Basque, and Welsh, thereby enabling consistent cross-country comparisons.

The {UK 2021 dataset} serves as the core of our analysis and includes tweets from MPs in the {UK national parliament} (London) as well as the {devolved parliaments of Scotland, Wales, and Northern Ireland}. Specifically, we collected {577 MPs’ tweets}, totaling {399,935} from the national parliament, {86,962} from Scotland, {42,787} from Wales, and {61,794} from Northern Ireland.

To assess the robustness and generalizability of our findings, we included parallel datasets from {Spain and Greece} for validation purposes. For Spain, we collected tweets from {279 MPs}, comprising {198,501} from the national parliament, {74,520} from Catalonia, and {15,324} from the Basque Country. For Greece, we gathered {76,655 tweets} from {184 MPs} in the national parliament.

These countries were selected due to their {diverse linguistic and socioeconomic contexts}, providing a valuable testbed for cross-national validation. The inclusion of the Spanish and Greek datasets enables us to examine whether the relationship between sentiment and engagement is consistent across languages and political environments.

%%%%%%%%%%%%%%%%%%%%%%%%%%%%%%%%%%%%%%%%%%%%%%%%%%%%%%%%%%%%%%%%%%%%%%%%%%%%%%%%%%%%%%%
\section{UNVEILING THE FEEDBACK LOOP}\label{sec:feedback}

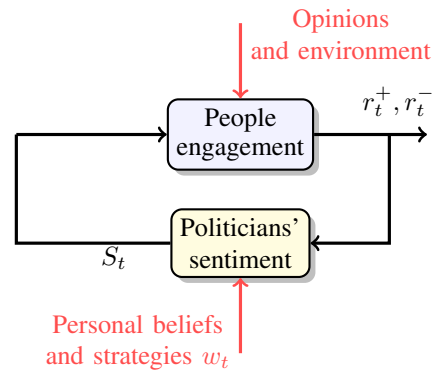
\begin{figure}[!ht]
  \centering
  \begin{tikzpicture}[node distance=2cm and 2cm, auto]
    \node (input) [coordinate] {};
    \node[draw, thick, fill=blue!5, rectangle, rounded corners, drop shadow, right=of input, align=center] (UpperBlock) 
    {\shortstack{People\\engagement}};
    \node[coordinate, above=1cm of UpperBlock] (incomingdata1) {};
    \node[coordinate, right=1cm of UpperBlock] (aid1) {};
    \node[draw, thick, fill=yellow!15, rectangle, rounded corners=0.9ex, drop shadow, below=0.5cm of UpperBlock, align=center] (LowerBlock) {Politicians'\\sentiment};
    \node[coordinate, below=1cm of LowerBlock] (incomingdata2) {};
    \node[coordinate, right=0.5cm of aid1] (output1) {};
    \node[coordinate, left=1cm of LowerBlock] (output2) {};
    \draw[->, very thick, red!70] (incomingdata2) -- node[near start, yshift=-3pt, align=center] {Personal beliefs\\and strategies $w_t$} (LowerBlock);
    \draw[->, very thick, red!70] (incomingdata1) -- node[near start, yshift=3pt, align=center] {Opinions\\and environment} (UpperBlock);
    \draw[->, very thick, black!60!black] (UpperBlock) -- node[near end, yshift=3pt] {$r^+_{t}, r^-_{t}$} (output1);
    \draw[->, very thick] (aid1) |- (LowerBlock);
    \draw[-, very thick] (LowerBlock) -| (input);
    \draw[->, very thick] (input) -- (UpperBlock);
    \draw[-, black!60!black] (LowerBlock) -- node[near end, yshift=2pt] {$S_{t}$} (output2);
  \end{tikzpicture}
  \caption{Schematic representation of the feedback mechanism in political communication.}
  \label{Fig:scheme}
\end{figure}

We investigate the hypothesis that political communication on social media is governed by a feedback mechanism: while public engagement is influenced by the sentiment of politicians' messages, {\it politicians in turn adapt the sentiment of their future communications to maximize engagement} (see Figure~\ref{Fig:scheme}).

To formalize this idea, we introduce an aggregate \textit{sentiment score} $S_t \in [-1, 1]$, defined for a group of MPs at time $t$ as the average sentiment of their tweets within that period. Each tweet is assigned a sentiment label of $+1$ (positive), $-1$ (negative), or $0$ (neutral), and $S_t$ is computed as the mean of these values. This score serves as a proxy for the overall emotional tone of political messaging at time $t$. Sentiment labels are obtained using a fine-tuned large language model (LLM), as described in the previous section.

We posit that the sentiment expressed at time $t+1$, namely $S_{t+1}$, is influenced by: 
\begin{enumerate}
\item[i.] the sentiment at time $t$, 
\item[ii.] the engagement metrics associated with the tweets posted at time $t$, and 
\item[iii.] internal factors such as personal beliefs or political strategy.
\end{enumerate} 
Formally, we hypothesize the following functional relationship:
\begin{equation}\label{eq:global_model}
	S_{t+1} = f(S_t, r^+_t, r^-_t, w_t),
\end{equation}
where $r^+_t$ and $r^-_t$ denote the fraction of retweets associated with positive and negative tweets, respectively, and $w_t$ denotes unobserved internal or strategic factors. Both $r^+_t$ and $r^-_t$ clearly fall within $[0, 1]$.

To empirically investigate this feedback mechanism, we adopt a linear predictor of the form:
\begin{equation}\label{eq:global_linear_model}
	\hat S_{t+1} = \alpha S_t + \beta r^+_t + \gamma r^-_t,
\end{equation}
where the coefficients $\alpha$, $\beta$, and $\gamma$ are learned from data, and $\hat S_{t+1}$ is the {\it predicted} sentiment at time $t+1$. Despite its simplicity, we show that this model captures key aspects of the sentiment adaptation process and provides interpretable insights into political behavior.

\begin{figure*}[h!]
	\centering
	\includegraphics[width=1\textwidth]{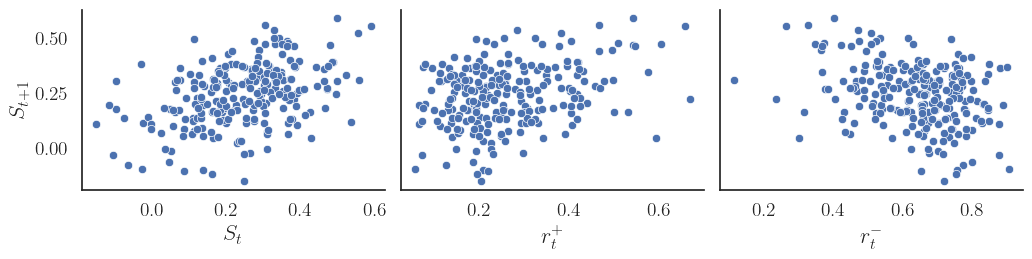}
	\caption{
	Pairwise scatter plots between future sentiment $S_{t+1}$ and the current variables: sentiment $S_t$, positive retweets $r^+_t$, and negative retweets $r^-_t$.
	}
	\label{fig:pairplot}
\end{figure*}
Figure~\ref{fig:pairplot} illustrates the pairwise relationships between $S_{t+1}$ and the input variables. The Pearson correlation coefficients suggest moderate linear dependencies: $0.42$ with $S_t$, $0.30$ with $r^+_t$, and $-0.32$ with $r^-_t$. Although with considerable variability, these results indicate that both sentiment and engagement metrics contribute to shaping future sentiment.

To rule out multicollinearity, we computed the Variance Inflation Factor (VIF) \cite{hair2006multivariate} for the regressors. The results are the following: VIF$(S_t) = 5.82$, VIF$(r^+_t) = 7.09$, and VIF$(r^-_t) = 2.97$. These values suggest that while there is some correlation among variables, it remains within acceptable bounds, and the contribution of each predictor is distinguishable.

\begin{figure}[!ht]
	\centering
	\includegraphics[width=1\columnwidth]{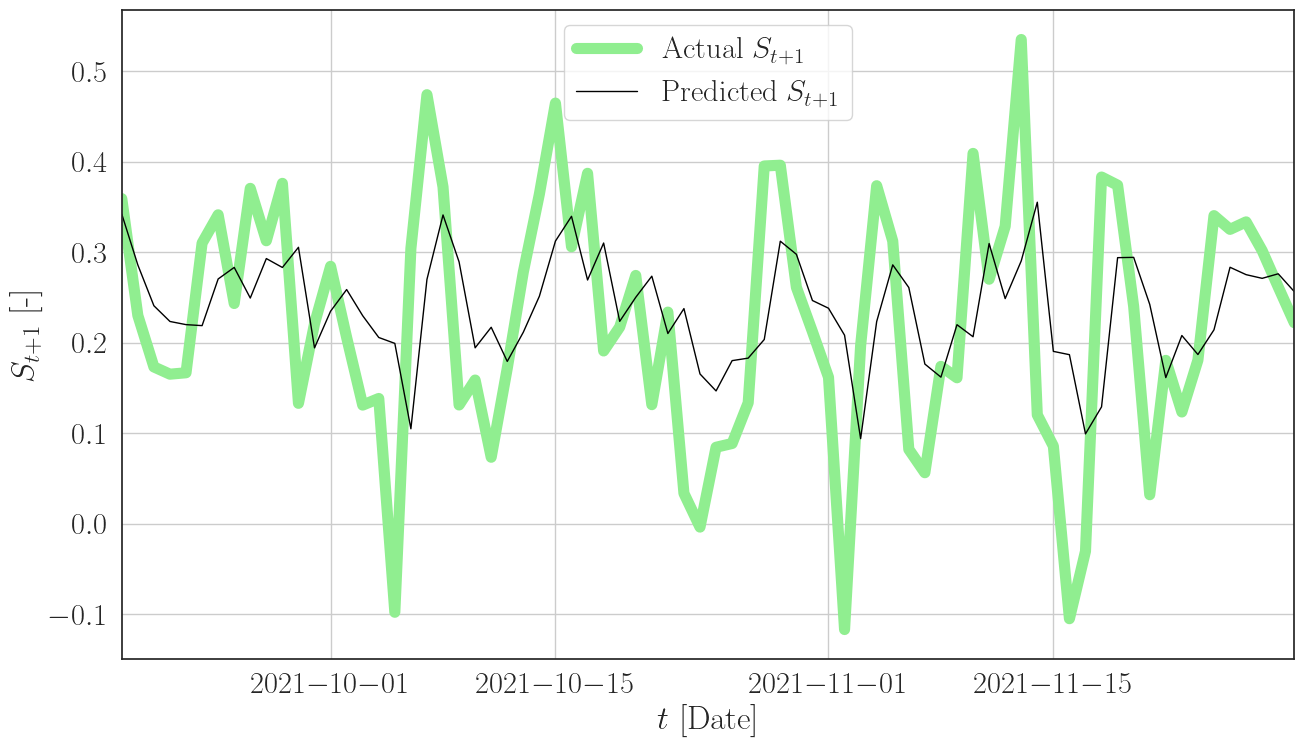}
	\caption{Prediction of future sentiment $S_{t+1}$ using current sentiment and engagement metrics. The model effectively captures sentiment trends despite variability.}
	\label{fig:global_prediction}
\end{figure}

The out-of-sample performance of the model in Equation~\eqref{eq:global_linear_model}, trained on 70\% of the UK 2021 dataset and tested on the remaining 30\%, is reported in Figure~\ref{fig:global_prediction}. The linear model captures the fluctuations in sentiment over the two-month test period, achieving a Root Mean Squared Error (RMSE) of $0.1419$. This outperforms a naive predictor that assumes $S_{t+1} = S_t$, which yields an RMSE of $0.1625$ (see Table~\ref{tab:rmse}).

\begin{table}[ht]
    \centering
    \caption{Root Mean Squared Error (RMSE) of Sentiment Prediction Models}
    \label{tab:rmse}
    \begin{tabular}{l|c}
        \toprule
        \textbf{Model} & \textbf{RMSE} \\
        \midrule
        Linear Feedback Model & 0.1419 \\
        Naive Baseline ($S_{t+1} = S_t$) & 0.1625 \\
        \bottomrule
    \end{tabular}
\end{table}

% %%%%%%%%%%%%%%%%%%%%%%%%%%%%%%%%%%%%%%%%%%%%%%%%%%%%%%%%%%%%%%%%%%%%%%%%%%%%%%%%%%%%
\section{MODELING INDIVIDUALS' SENTIMENT DYNAMICS}\label{sec:individuals}

While the previous section analyzed the sentiment-feedback loop from a global perspective, we now investigate how individual politicians respond to engagement signals. Specifically, we define $S^i_t$ as the sentiment score of politician $i$ at time $t$ and consider the dynamics:
\begin{equation}\label{eq:local_model}
	S^i_{t+1} = f(S^i_t, r^+_t, r^-_t, w_t),
\end{equation}
where $r^+_t$ and $r^-_t$ are still the fraction of total positive and negative retweets at time $t$, namely they are still {\it global} engagement signals, whereas $w_t$ captures individual-level factors such as personal beliefs or strategic decisions.

We approximate this dynamic using the following linear predictor:
\begin{equation}\label{eq:local_linear_model}
	\hat S^i_{t+1} = \alpha_i S^i_t + \beta_i r^+_t + \gamma_i r^-_t,
\end{equation}
where the coefficients $\alpha_i$, $\beta_i$, and $\gamma_i$ are estimated from data and represent the {\it sensitivity} of each politician’s future sentiment to their current sentiment and to engagement signals.

Note that the global sentiment score $S_t$ previously introduced can be interpreted as an aggregate of individual behaviors, while the feedback loop operates at the individual level on $S^i_{t}$. In practice, politicians observe and internalize the aggregated sentiment emerging from users’ reactions to their posts, such as the overall tone of comments or engagement trends, which they interpret as a signal of public approval or disapproval, and consequently adapt their future communication strategies.

This disaggregated modeling enables us to assess, for each politician, {\it whether and how} engagement influences sentiment adjustment. The coefficient $\alpha_i$ captures sentiment inertia: if $\alpha_i > 0$, the politician tends to maintain their current sentiment; if $\alpha_i < 0$, the politician tends to reverse it. The coefficients $\beta_i$ and $\gamma_i$ indicate the influence of positive and negative retweets, respectively. If $\beta_i > \gamma_i$, the politician appears more responsive to positive feedback; if $\beta_i < \gamma_i$, they are more affected by negative engagement.

To assess systematic differences across political alignments, we focus on highly active users, namely those tweeting on more than 90\% of days and thus providing a dataset rich enough for modeling. Then, let us define the difference variable $\delta_i = \beta_i - \gamma_i$ for each MP. We compute its {\it standardized z-score} \cite{hair2006multivariate} as:
\begin{equation}
	z_i = \frac{\delta_i - \mu_{\delta}}{\sigma_{\delta}},
\end{equation}
where $\mu_{\delta}$ and $\sigma_{\delta}$ are the sample mean and standard deviation of $\delta_i$ across all considered politicians.

\begin{figure}[!ht]
	\centering
	\includegraphics[width=1\columnwidth]{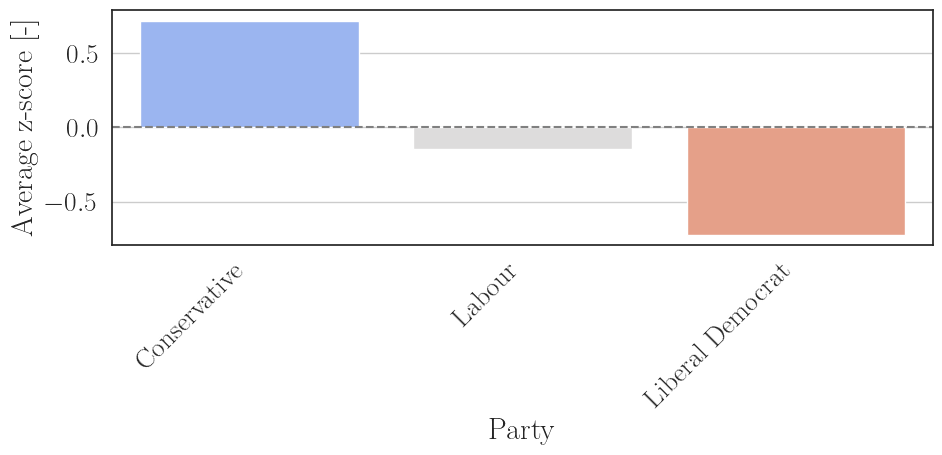}
	\caption{Z-scores of the difference between positive and negative retweet coefficients, grouped by party (UK 2021 dataset). Positive scores indicate greater influence of positive feedback; negative scores, greater influence of negative feedback.}
	\label{fig:z-score_uk}
\end{figure}

Figure~\ref{fig:z-score_uk} reports the distribution of $z_i$ scores by party for the UK. Positive values indicate a stronger sensitivity to positive feedback, while negative values suggest a greater influence of negative engagement. Conservative MPs exhibit predominantly positive z-scores, consistent with a communication style oriented toward optimism and affirmation. In contrast, MPs from Labour and the Liberal Democrats show negative z-scores, reflecting the more critical tone of that period. The difference in z-score magnitude between these two opposition parties may stem from sample size disparities (125 for Liberal Democrats vs. 1,894 for Labour), which affect statistical robustness.

\begin{figure}[!ht]
	\centering
	\includegraphics[width=1\columnwidth]{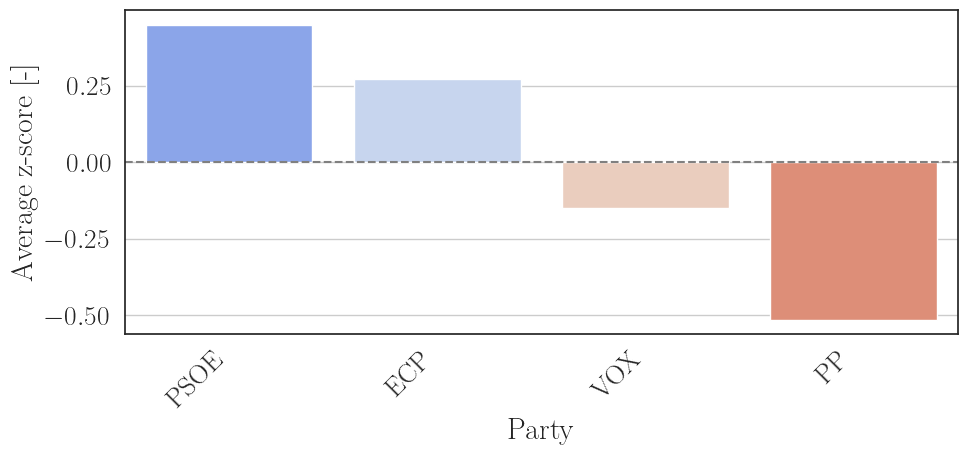}
	\caption{Z-scores of the difference between positive and negative retweet coefficients, grouped by party (Spain 2021 dataset).}
	\label{fig:z-score_es}
\end{figure}

\begin{figure}[!ht]
	\centering
	\includegraphics[width=1\columnwidth]{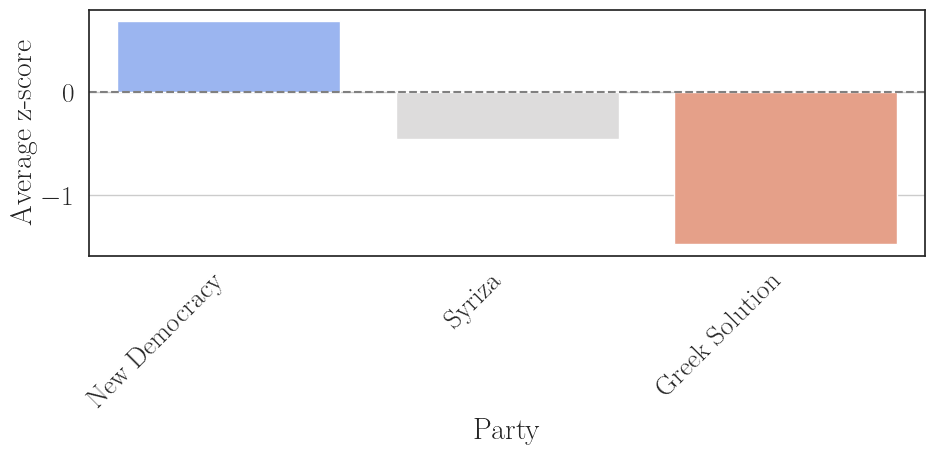}
	\caption{Z-scores of the difference between positive and negative retweet coefficients, grouped by party (Greece 2021 dataset). Only parties with at least 100 active users are included.}
	\label{fig:z-score_gr}
\end{figure}

For completeness, we present analogous results for Spain and Greece in Figures~\ref{fig:z-score_es} and~\ref{fig:z-score_gr}, respectively. In the Greek case, only parties with at least 100 active users were included to ensure statistical reliability.

In 2021, Spain was governed by Pedro Sánchez of the Spanish Socialist Workers’ Party (PSOE). The positive z-scores for PSOE (and its coalition partner ECP) suggest a preference for positive messaging, in line with Sánchez’s communication style. In contrast, the opposition parties - People’s Party (PP) and Vox - display more negative z-scores, consistent with their more confrontational tone and the significant rise of Vox during that period.

Greece was led by the New Democracy (ND) party, with Prime Minister Kyriakos Mitsotakis presiding over a parliamentary majority. The primary opposition party was Syriza, headed by Alexis Tsipras. Our analysis reveals that ND politicians tend to exhibit higher z-scores, indicating more positive communication, whereas Syriza and nationalist parties like Greek Solution — known for their critical and anti-EU stances — show more negative values, aligned with their oppositional roles and rhetoric.

There appear to be no systematic or statistically significant differences in the value of the ``inertia'' coefficient $\alpha_i$ across parties, suggesting that sentiment persistence is a common feature of political communication regardless of party affiliation. The coefficient is consistently positive and significantly greater than zero, indicating a strong tendency of all politicians in the sample to maintain their prevailing sentiment over time. In the United Kingdom and Spain, $\alpha_i$ is slightly higher for government parties, although this pattern does not hold in Greece. However, the magnitude of these differences is not statistically significant.

%%%%%%%%%%%%%%%%%%%%%%%%%%%%%%%%%%%%%%%%%%%%%%%%%%%%%%%%%%%%%%%%%%%%%%%%%%%%%%%%%%%%%%%
\section{THEORETICAL ANALYSIS OF THE CLOSED-LOOP SYSTEM}\label{sec:theory}

To complement the empirical evidence presented in previous sections, we now turn to the theoretical behavior of the sentiment–engagement system under full feedback closure. Our aim is to study how sentiment and engagement co-evolve when politicians adapt their expressed tone in response to engagement, and engagement itself is simultaneously shaped by sentiment.

\subsection{Linear regime}
We begin with the case where the dynamics of average sentiment $S_t$ follow the same linear model used in predictive form in \eqref{eq:global_linear_model}, namely
\begin{equation}\label{eq:global_linear_model_recall}
   S_{t+1} = \alpha S_t + \beta\, r^+_t + \gamma\, r^-_t.
\end{equation}
To close the loop, we introduce a simple mapping from sentiment to engagement that ensures non-negative fractions:
\begin{subequations}\label{eq:engagement}
  \begin{align}
r^+_{t+1} = a \cdot \frac{1 + S_t}{2}, \\
r^-_{t+1} = b \cdot \frac{1 - S_t}{2},
  \end{align}
\end{subequations}
with $a,b \ge 0$ scaling the maximal levels of positive and negative engagement. Selecting the mapping like this implies:
\begin{itemize}
    \item $S_t = 1$: full positive engagement, $r^+_{t+1} = a$, $r^-_{t+1} = 0$;
    \item $S_t = -1$: full negative engagement, $r^+_{t+1} = 0$, $r^-_{t+1} = b$;
    \item $S_t \in (-1,1)$: mixed engagement, with both $r^+_{t+1}$ and $r^-_{t+1}$ strictly positive.
\end{itemize}
Thus, $r^+_{t+1}$ increases monotonically with $S_t$, while $r^-_{t+1}$ decreases, reflecting the intuition that positive tone generates positive reactions and vice versa.

Substituting \eqref{eq:engagement} into \eqref{eq:global_linear_model_recall} yields a linear second-order difference equation:
\begin{equation}\label{eq:closed_loop}
S_{t+2} - \alpha S_{t+1} - \frac{\beta a - \gamma b}{2} S_t = \frac{\beta a + \gamma b}{2}.
\end{equation}
Here, $\alpha$ measures sentiment inertia: higher values capture politicians’ tendency to maintain past tone. The term $(\beta a - \gamma b)/2$ represents the net effect of engagement on future sentiment: positive values amplify sentiment, while negative values attenuate it. 

\begin{figure}[h!]
    \centering
    \includegraphics[width=1\columnwidth]{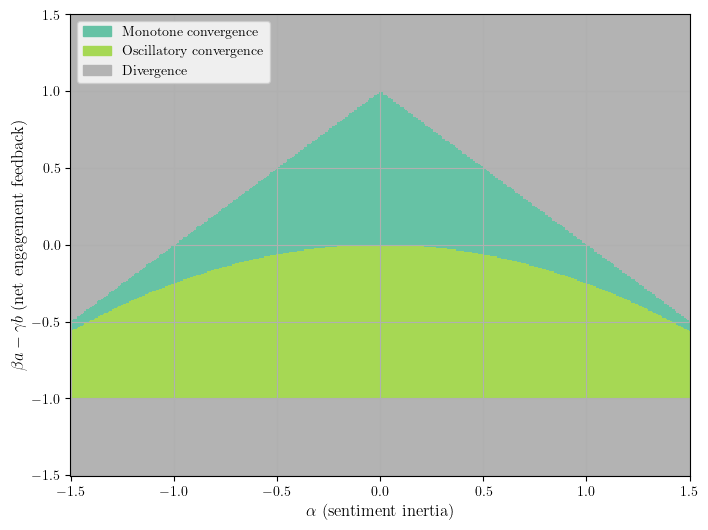}
    \caption{Stability diagram of the closed-loop linear sentiment system. Green: monotone convergence; light green: damped oscillatory convergence; grey: divergence.}
    \label{fig:stability_diagram}
\end{figure}

Figure~\ref{fig:stability_diagram} illustrates the qualitative stability regimes of the system across the $(\alpha,\, \beta a - \gamma b)$ parameter plane:
\begin{itemize}
    \item \textbf{Monotone convergence:} sentiment smoothly approaches a steady state.
    \item \textbf{Oscillatory convergence:} sentiment exhibits damped swings before settling.
    \item \textbf{Divergence:} feedback is too strong, leading to instability.
\end{itemize}

Politically, this means that moderate engagement feedback combined with realistic sentiment inertia keeps aggregate discourse stable. In contrast, excessive feedback risks amplifying swings and destabilizing public debate. Empirical estimates for our dataset fall in the monotone convergence region, consistent with the observed stability of online discussions even in this simplified (linear) setting.

\subsection{Nonlinear regime}
As $S_t$ approaches $\pm 1$, engagement fractions saturate and the linear description \eqref{eq:closed_loop} no longer holds. To model this explicitly, we introduce a clipping operator
\[
\operatorname{sat}_{[-1,1]}(x) := \min\{\max\{x,-1\},1\},
\]
and write the saturated dynamics as
\begin{equation}\label{eq:closed_loop_saturated}
S_{t+1} = \operatorname{sat}_{[-1,1]}\!\left(\alpha S_t + \beta\,a\frac{1+S_{t-1}}{2} + \gamma\,b\frac{1-S_{t-1}}{2}\right).
\end{equation}
This formulation highlights two nonlinearities: (i) explicit saturation of $S_t$ within $[-1,1]$, and (ii) affine dependence on the past state $S_{t-1}$. Together, these make the political feedback loop a \emph{piecewise-affine}, second-order map.

Saturation introduces several key features:

\paragraph{Invariant set and boundedness.}
The interval $[-1,1]$ is forward-invariant: trajectories remain bounded regardless of coefficients. This reflects the realistic cap imposed by social attention (public sentiment cannot grow unboundedly positive or negative).

\paragraph{Fixed points and local linearization.}
If an equilibrium $S^\star \in [-1,1]$ satisfies the unsaturated fixed-point equation, then local dynamics are governed by the linearized system, and stability follows from the characteristic polynomial. Hence, the linear analysis applies locally to interior equilibria.

\paragraph{Boundary equilibria and locking.}
If the unconstrained equilibrium lies outside $[-1,1]$, the system admits a \emph{boundary equilibrium} at $S^\star=\pm1$, where saturation is active. Dynamics then reduce to a clipped, lower-dimensional map. Politically, this corresponds to \emph{locked-in discourse}: once public sentiment reaches an extreme, engagement cannot push it further, and the system may remain stuck until disrupted by external shocks.

\paragraph{Oscillations vs. damping.}
The net feedback coefficient $(\beta a-\gamma b)/2$ governs local oscillatory behavior:
\begin{itemize}
  \item Moderately negative values (greater than $-1$) yield damped oscillations around an interior equilibrium.
  \item Strongly negative values can sustain oscillations or produce high-period cycles through border-collisions at the boundaries.
  \item Large positive values lead to rapid divergence in the unsaturated model; saturation then clips sentiment to $\pm1$, producing \emph{plateauing} at the extremes.
\end{itemize}

Politically, these regimes suggest that moderate counter-phase responses generate short-lived swings that stabilize. Conversely, strong opposing feedback may sustain cycles of alternating tone, while strong reinforcing feedback drives discourse rapidly toward entrenched extremes.

% %%%%%%%%%%%%%%%%%%%%%%%%%%%%%%%%%%%%%%%%%%%%%%%%%%%%%%%%%%%%%%%%%%%%%%%%%%%%%%%%%%

\section{DISCUSSION AND CONCLUSIONS}\label{sec:conclusions}

This work aimed to delve deep into the intricate feedback dynamics governing political communication on social media. Specifically, the current study provides a crucial behavioral explanation for the findings of \cite{antypas2023negativity}. It suggests that when a politician's negative tweets receive high engagement, their subsequent tweets tend to become less positive, or more negative. This closes the feedback loop: higher virality of negative content (known from \cite{antypas2023negativity}) reinforces the tendency of politicians to use more negative rhetoric (as shown in this paper), thereby perpetuating the spread of negativity.

Moreover, the observed differences in individual politician's sentiment dynamics - expressed by their model parameters - directly align with the party-level sentiment distributions identified in \cite{antypas2023negativity}. More specifically: 
\begin{itemize} 
	\item The fact that \textbf{opposition politicians} exhibit greater sensitivity to negative feedback aligns with their general tendency to post more negatively charged tweets. This can be interpreted as a strategic response: if critical or negative messages resonate more with the public, opposition parties might lean into this to garner greater engagement and challenge governing parties' positions.
	\item \textbf{Governing party politicians} instead show greater sensitivity to positive feedback, consistently with their overall more positive communication. They aim to project a positive image of the country's state and their agenda. 
\end{itemize}
In conclusion, while \cite{antypas2023negativity} empirically demonstrates that negativity spreads faster in political communication, this paper provides evidence for how this phenomenon is {sustained}: through politicians' adaptive communication strategies driven by engagement signals. This creates a reinforcing cycle that contributes to the increasingly polarized and negative tone observed in online political discourse. 

As a final contribution, this paper has outlined hypothetical scenarios that may emerge when the feedback loop is fully closed, offering a theoretical framework to understand the stability and evolution of political sentiment over time, even under parameter values different from those observed in the analyzed dataset.

\subsection{Possible future research directions}

Interesting extensions include the temporal and geographic expansion of the analysis to a wider range of time periods, and to additional countries or political systems. Further work should focus on model refinement and complexity, and investigating the nature and influence of unobserved internal or strategic factors ($w_t$) that shape politicians' sentiment. 
Implementing a more fine-grained sentiment analysis, such as a 1-5 scale or considering specific emotional aspects beyond simple polarity, could reveal subtler relationships between content and virality. 

As a final remark, we believe that the robust methodology developed in this research could be broadly leveraged for subsequent large-scale sociopolitical studies (e.g., analyzing public reactions to climate change or natural disasters) and could serve as a valuable ``dashboard'' to monitor and potentially inform interventions aimed at mitigating negativity in online political discourse.

%%%%%%%%%%%%%%%%%%%%%%%%%%%%%%%%%%%%%%%%%%%%%%%%%%%%%%%%%%%%%%%%%%%%%%%%%%%%%%%%
\section*{ACKNOWLEDGMENTS}

We thank the authors of \cite{antypas2023negativity} for sharing their dataset and providing valuable insights into the virality of political messages. Their work laid the foundation for this analysis and interpretation of feedback dynamics in political communication. 

We are also deeply grateful to Prof. Giulio Panzani and Prof. Valentina Breschi for their careful reading of the draft and for their insightful comments, which greatly improved the clarity of this work.

%%%%%%%%%%%%%%%%%%%%%%%%%%%%%%%%%%%%%%%%%%%%%%%%%%%%%%%%%%%%%%%%%%%%%%%%%%%%%%%%

\bibliographystyle{IEEEtran}
\bibliography{bibliography}

\end{document}